\documentclass[12pt]{iopart}

\usepackage{graphicx}

\def\mbi#1{\mbox{\boldmath$#1$}}
\def\kp{\mbi{k} \cdot \mbi{p}}

\def\beeq{\begin{equation}}
\def\eneq{\end{equation}}
\def\beeqa{\begin{eqnarray}}
\def\eneqa{\end{eqnarray}}

\begin{document}

\title{Novel electronic wave interference patterns in nanographene sheets}
\author{Kikuo Harigaya\dag \ddag \S
\footnote[4]{To
whom correspondence should be addressed
(E-mail: {\tt k.harigaya@aist.go.jp},
URL: {\tt http://staff.aist.go.jp/k.harigaya/})},
Yousuke Kobayashi\S, Kazuyuki Takai\S, J\'{e}r\^{o}me Ravier\S, 
and Toshiaki Enoki\S}
\address{
\dag Nanotechnology Research Institute, AIST, Tsukuba 305-8568, Japan\\
\ddag Synthetic Nano-Function Materials Project, AIST, 
Tsukuba 305-8568, Japan\\
\S Tokyo Institute of Technology, Oh-okayama, Meguro-ku 152-8551, Japan}
\begin{abstract}
Superperiodic patterns with a long distance in a nanographene 
sheet observed by STM are discussed in terms of the interference 
of electronic wave functions.  The period and the amplitude of 
the oscillations decrease spatially in one direction.  We explain 
the superperiodic patterns with a static linear potential theoretically.  
In the $\kp$ model, the oscillation period decreases, and agrees with
experiments. The spatial difference of the static potential is 
estimated as 1.3 eV for 200 nm in distance, and this value seems 
to be natural. It turns out that the long-distance oscillations 
come from the band structure of the two-dimensional graphene sheet.\\

\noindent
PACS: 73.61.TM, 73.20.At, 71.10.-w
\end{abstract}

\section{Introduction}

Nanographene sheets are attractive materials, because their 
magnetic and transport properties show novel and peculiar 
properties, originating from nonbonding states localized at 
the zigzag-edges [1,2]. Theoretical works [2-5] have been 
performed to clarify the mechanisms of the unique magnetism. 
It has been found that the A-B stacking and the presence 
of the localized electronic spins originating in the open 
shell nature are the favorable conditions for the magnetism.

On the other hand, direct observation by scanning tunneling 
microscope (STM) is powerful for structural analysis as 
well as for investigation of electronic properties. Previously, 
we have reported that the interlayer distance of a single 
nanographene on a highly oriented pyrolytic graphite (HOPG) 
substrate is about 0.35-0.37 nm and is larger than that of 
bulk material [6]. In recent study, we have found 
superperiodic patterns with extremely long periodicity in 
the STM images of a nanographene sheet, which varies 
spatially [7]. In this paper, we give an explanation for 
this novel results in terms of the interference of 
electronic wave functions.

In order to understand the origins of the long-distance 
oscillations which were observed, we will make a comparison with 
the theoretical electron densities using the free electron model 
confined within an infinite square well, and also using the 
$\kp$ model [8,9] for two-dimensional graphene sheet.  One of 
the present authors has used the model for the understandings 
of the multi channel Kondo effect [10] and the Cooper pair 
propagation [11] in metallic carbon nanotubes. In the graphene 
sheets, superperiodic patterns in STM images due to the moir\'{e} 
origins for the A-B stacking [12] and structural deformations [13] 
have been reported in the literatures. However, the observed 
superperiodic patterns with quite long periods over 10 nm 
seem not coming from the moir\'{e} mechanism, and our finding 
calls for new interpretations. We assume the presence of a 
static potential with a linear decrease in one direction. The 
calculated local electron density will be compared with the 
experiments. We will clarify that the long-distance 
oscillations are due to the presence of electrons with the 
band structure of a two-dimensional graphene sheet.

\begin{figure}
\begin{center}
\includegraphics{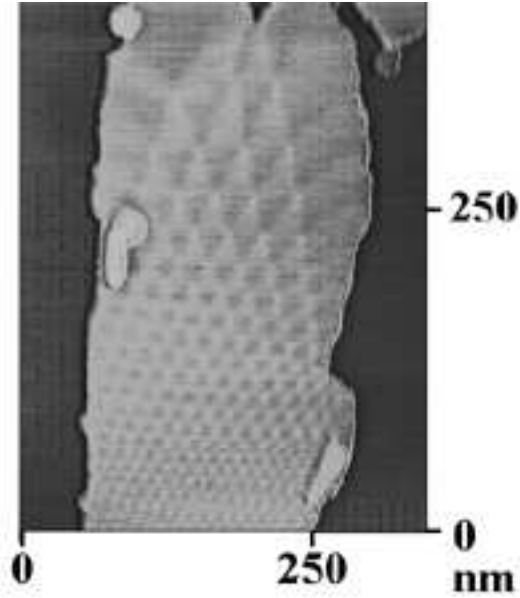}
\end{center}
\caption{
STM image of the superperiodic pattern on a necktie-shaped
graphene plate on HOPG substrate.}
\label{Fig1}
\end{figure}

This paper is organized as follows. In section 2, the experimental 
results are briefly reviewed. In section 3, calculations by the 
free electron model are compared with experiments. In section 4, 
the analysis with the $\kp$ model is performed. The paper is 
closed with summary in section 5.

\section{STM observations}

In this section, the experimental data is briefly reviewed. 
In Fig. 1, a STM image of the graphene sheet with a necktie 
shape is shown.  The detail will be published elsewhere.  The 
observation has been done with the following condition: bias 
voltage V = 200 mV and current I = 0.7 nA. The distance between 
the graphene necktie and the substrate is over 0.8 nm, 
suggesting that it consists of a stacking of two graphene layers,  
which interact weekly with the HOPG substrate. Interestingly, 
the period and the amplitude of the oscillations decrease from 
the top to the bottom along the graphene necktie. The oscillation 
period is one order of magnitude larger than that of the moir\'{e} 
pattern due to stacking [12], and therefore this possibility 
can be excluded. We can assume effects of long-distance 
periodic-structural deformations [13] in the graphene surface 
or interference effects of electronic wave functions.

We have also observed that the oscillation period becomes 
longer by placing a nanographene flake on the graphene necktie, 
as shown in Fig. 2. The oscillations period seems to be double 
in the upper region of the necktie after addition of one flake. 
The oscillation below the flake seems to be only slightly 
modified by the flake. Such effect on the oscillations cannot 
be explained by some structural modulations. Therefore, the 
oscillation patterns could be the effect of interference of 
the electronic wave functions in the graphene surface.

\begin{figure}
\begin{center}
\includegraphics{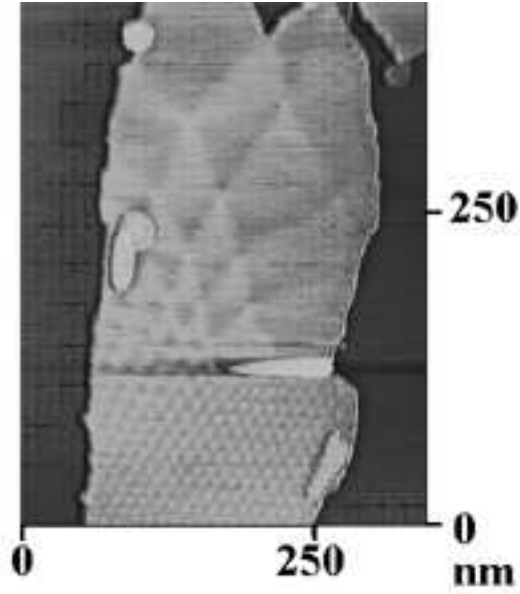}
\end{center}
\caption{
STM image of the superperoidic pattern on the necktie-shaped
graphene sheet observed in Fig. 1 after a nanographene flake 
is placed.}
\label{Fig2}
\end{figure}

\section{Free electron model}

We will characterize the interference patterns theoretically.
Two dimensional coordinate is defined so that the $y$-axis
is along with the long direction of the graphene necktie.
The $x$-axis is perpendicular to the $y$-axis.
By assuming the electric static potential $-F \cdot y$ 
which is proportional to the $y$-axis, and the 
confinement effect due to the well-shaped potential 
within $-d/2 < x < d/2$, we obtain the Schr\"{o}dinger
equation:
\beeq
\left[-\frac{\hbar^2}{2m} 
\left( \frac{\partial^2}{\partial x^2} + 
\frac{\partial^2}{\partial y^2} \right) 
+ V_{\rm well}(x) - F \cdot y \right]
\psi(x,y) = E \psi(x,y),
\eneq
where $V_{\rm well} (x)$ means the well potential with the
infinite depth.  The electron density with the energy $E$
is written as:
\beeq
|\psi (x,y)|^2 = \sum_n a_n | \psi_x (E_n) \psi_y(E-E_n)|^2,
\eneq
where $a_n$ is the coefficient of the occupancy with the
quantum number $n$, $\psi_x(E_n)$ is the solution in the 
well in the $x$-direction, and $\psi_y(E-E_n)$ is the
solution for the potential term $-F \cdot y$.  The solution
in the well is: 
\beeq
\psi_x(E_n) = 
\left\{ \begin{array}{c}
A {\rm cos} \left(\frac{n\pi x}{d} \right) \mbox{for odd $n$}\\
A {\rm sin} \left(\frac{n\pi x}{d} \right) \mbox{for even $n$}
\end{array} \right. ,
\eneq
with $E_n = \pi^2 \hbar^2 n^2 / 2 m d^2$ and $A = \sqrt{2/d}$.
The solution for the linear potential is:
\beeq
\psi_y(E-E_n) = \Phi \left[- \left( \frac{2mF}{\hbar^2} \right)^{1/3}
\left( x + \frac{E}{F} - \frac{E_n}{F} \right) \right],
\eneq
where $\Phi(x)$ is the Airy function
\beeq
\Phi(x) = \frac{1}{\sqrt{\pi}} \int_0^\infty
{\rm cos} \left( \frac{u^3}{3} + ux \right) {\rm d}u.
\eneq
The local density of states of electrons with $a_n = 1$  
only for $n=4$ is shown in Fig. 3.  We can theoretically 
explain the decrease of the oscillation period and the 
amplitude along the $y$-direction.  This property is owing
to the form of the Airy function.  There is a standing 
wave in the $x$-direction.  However, we cannot explain the 
details of oscillations in the $x$-direction of Fig. 1, 
possibly by the effects of the complex shapes of the 
boundary in the graphene necktie.

\begin{figure}
\begin{center}
\includegraphics{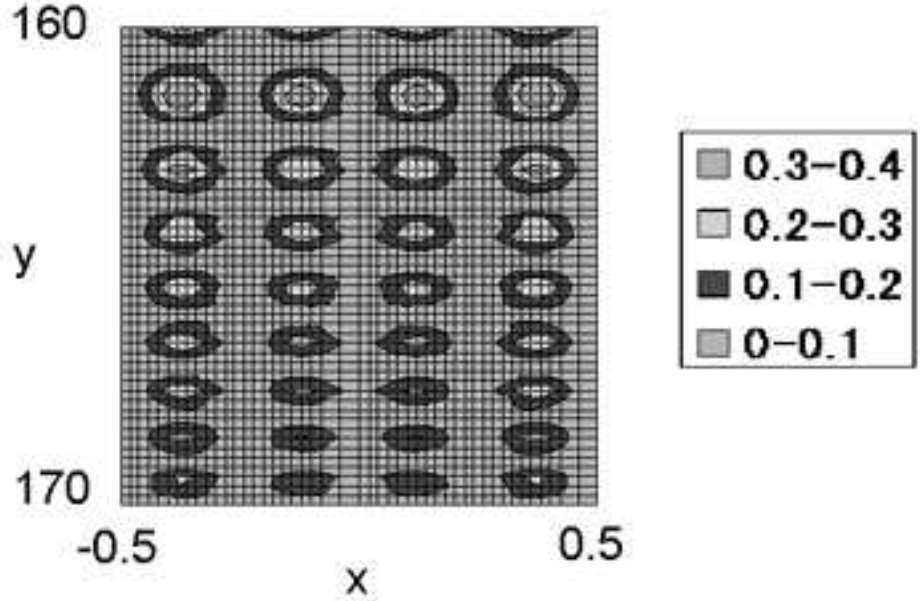}
\end{center}
\caption{
Two dimensional plot of the electron density
calculated by the free electron model.  The bottom and
left axes are shown in arbitrary units.  The quantum number
$n=4$ is taken for the standing wave within the infinite well, 
$-0.5 < x < 0.5$ with $d = 1$.}
\label{Fig3}
\end{figure}

Even though the oscillation in the well is uniform spatially,
we can compare the oscillation patterns in the $y$-direction
at least.  Figure 4 shows comparison with the experiment
where the peak positions along the long axis of the necktie
are plotted.  The decrease of the amplitude of the theoretical 
curve seems more rapid than that of the experimental data.
The strength of the static potential is $F = 5.26 \times
10^{-6}$ eV/nm with using the free electron mass.  The potential 
variation over the distance 200 nm is $1.1 \times 10^{-3}$ eV, 
and this is quite small. Phonon fluctuation effects or the
presence of impurities can override such the small potential
change.  This difficulty might be due to the assumption of 
the free electron model of this section.

\begin{figure}
\begin{center}
\includegraphics{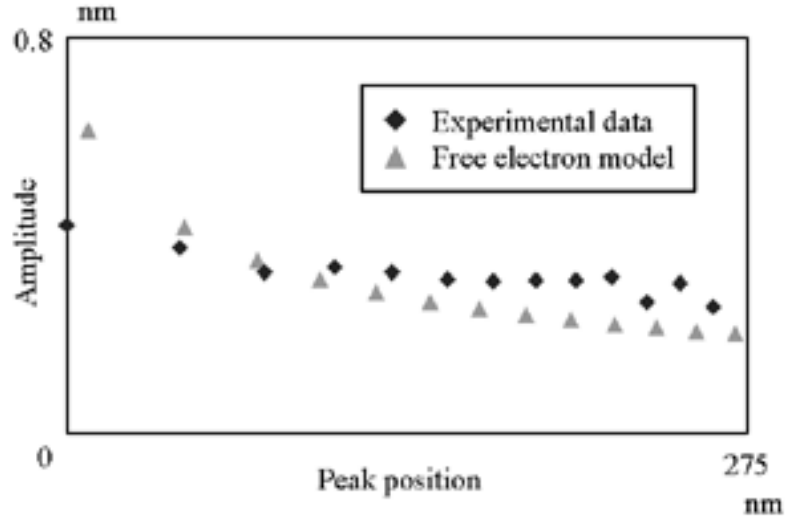}
\end{center}
\caption{
Comparison for the electron wave patterns by STM 
and the free electron theory.  Experimental peak
positions along the perpendicular direction of Fig. 1 are
plotted by diamonds.  The results of the fitting by
one dimensional free electron model are shown by triangles.}
\label{Fig4}
\end{figure}

\section{Continuum $\kp$ model}

In order to substantiate analysis of the interference patterns,
we give comparison with the calculation of the model for
the graphene plane.  Here, we use the continuum $\kp$ model [7,8].
The hamiltonian around the K point with the linear potential 
$- F \cdot y$ is:
\beeq
H =
\left( \begin{array}{cc}
- F \cdot y & -i \gamma \frac{\partial}{\partial x} 
- \gamma \frac{\partial}{\partial y}\\
-i \gamma \frac{\partial}{\partial x} 
+ \gamma \frac{\partial}{\partial y} & - F \cdot y
\end{array} \right),
\eneq
where $\gamma \equiv (\sqrt{3}/2) a \gamma_0$, $a$ is the 
bond length, and $\gamma_0$ is the hopping integral between
neighboring carbon atoms.  This model is solved together with
the infinite well potential $V_{\rm well} (x)$ as used in 
the previous section.  The Schr\"{o}dinger equation 
$H \Psi =E \Psi$ gives a oscillating solution:
\beeq
\Psi = 2A
\left( \begin{array}{c}
{\rm sin} \left( \frac{E_n x}{\gamma} \right) \\
-i {\rm cos} \left( \frac{E_n x}{\gamma} \right)
\end{array} \right) \otimes
\left( \begin{array}{c}
{\rm sin} \left[ \frac{1}{\gamma} 
\left( \tilde{E}y+\frac{1}{2}Fy^2 \right) \right] \\
{\rm cos} \left[ \frac{1}{\gamma} 
\left( \tilde{E}y+\frac{1}{2}Fy^2 \right) \right]
\end{array} \right),
\eneq
where $E_n = n\pi \gamma /d$ and $\tilde{E} = E - E_n$.
The electron density at the A-sublattice point is calculated as:
\beeqa
|\psi_{\rm A}(\mbi{R}_{\rm A})|^2
&=& 4A^2 {\rm sin}^2 \left( \frac{n \pi x}{d} \right) \nonumber \\
&\times& \left\{ 1+{\rm cos} [(\mbi{K}-\mbi{K'})\cdot \mbi{R}_{\rm A} ]
{\rm sin} \left[ \frac{2}{\gamma} 
\left( \tilde{E}y+\frac{1}{2}Fy^2 \right) \right] \right\}
\eneqa
where $\mbi{R}_{\rm A}$ is the lattice point of the A-sublattice,
$\mbi{K}$ and $\mbi{K'}$ are the K and K' points in the wave
number space.  We pay attention to the long period oscillating component:
\beeq
{\rm sin}^2 \left(\frac{n \pi x}{d} \right)
\left[ {\rm const.} + {\rm sin}
\left( \frac{F y^2}{\gamma} - \frac{2 n \pi}{d}y \right) \right],
\eneq
where we take $E=0$ at the Fermi energy.  This functional
form for the quantum number $n=4$ is plotted in Fig. 5 with 
the assumption $d=1$.  The amplitude is spatially constant, 
and the oscillation period becomes smaller as $y$ becomes 
larger.  We can explain the decrease of the oscillation period 
found in experiments of Fig. 1, though the uniform array of the
standing wave would be the result of the simplified
theory and this is in contrast with the observations.

\begin{figure}
\begin{center}
\includegraphics{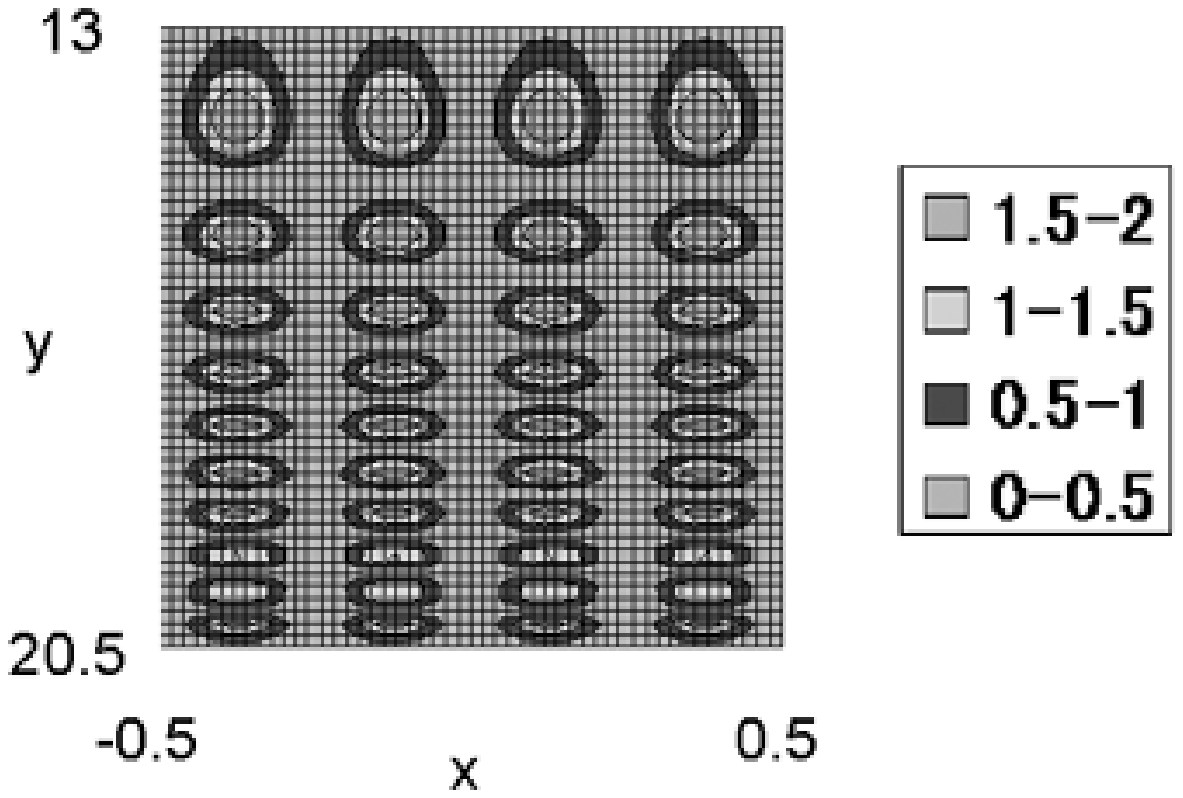}
\end{center}
\caption{
Two dimensional plot of the electron density
calculated by the $\kp$ model.  The bottom and
left axes are shown in arbitrary units.  The quantum number
$n=4$ is taken for the standing wave within the infinite well, 
$-0.5 < x < 0.5$ with $d = 1$.}
\label{Fig5}
\end{figure}

The peak positions of the electron density in the long direction
of the graphene necktie of Fig. 1 are plotted in Fig. 6, and
comparison with the result of eq. (9) is given.  The slight
decrease found in the experiments cannot be reproduced by
the result of the $\kp$ model.  However, the decrease of the 
oscillation period fairly agrees with the experiments.
The fitting gives the parameter of the potential gradient 
$F=6.49 \times 10^{-3}$ eV/nm.
The total potential variation over the distance 200 nm
becomes 1.3 eV.  Such magnitude of the potential change
would survive thermal lattice fluctuations and can really
exist in experiments.  The present result by no means implies that
the wave functions observed with superperiodic amplitudes
are of the electrons which have energy levels of the
graphene plane.

\begin{figure}
\begin{center}
\includegraphics{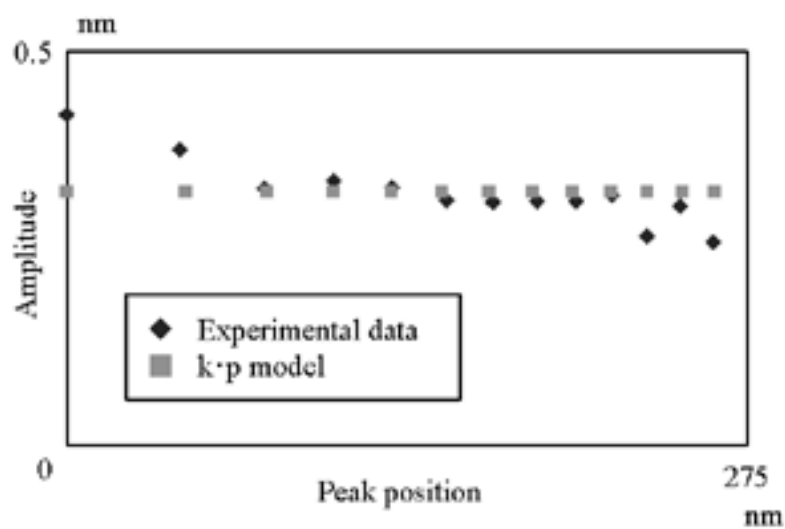}
\end{center}
\caption{
Comparison for the electron wave patterns by STM 
and the $\kp$ model.  Experimental peak
positions along the perpendicular direction of Fig. 1 are
plotted by diamonds.  The results of the fitting by
the long distance envelope functional form derived from
the $\kp$ model are shown by squares.}
\label{Fig6}
\end{figure}

\section{Summary}

Superperiodic patterns in a nanographene sheet observed by STM 
can be explained with two models of electronic wave functions 
in terms of the interference. First, the experimental results 
have been briefly introduced.  The period and the amplitude of the 
oscillations decrease spatially in one direction along with 
the long direction of the graphene necktie.  The patterns are 
superperiodic, and the period is one order of the magnitudes 
longer than that of the well known moir\'{e} pattern observed 
in A-B stacked graphite systems.  This is a novel finding in 
our experiments.

Next, theoretical characterizations have been reported.  We 
have explained the interference patterns with the static linear 
potential by using a free electron model and also by the 
continuum $\kp$ model.  In the free electron model, we have
derived the decrease of the oscillation period and the amplitude 
along the decreasing direction of the linear potential.  However, 
the strength of the linear potential turned out to be 
unrealistically too small.  In the $\kp$ case, the oscillation 
period decreases, and the amplitude is constant.  The spatial 
difference of the static potential is estimated as 1.3 eV for 
the distance 200 nm, and this value seems to be natural.  
It turned out that the long distance oscillations 
come from electrons with the band structures of the two dimensional
graphene sheet.

\section*{Acknowledgements}
Useful discussion with members of the Nanomaterials Theory Group,
Nanotechnology Research Institute ({\tt http://unit.aist.go.jp/nanotech/}), 
AIST and the Enoki-Fukui Laboratory, Department of Chemistry, 
Tokyo Institute of Technology is acknowledged.
This work has been partly supported by NEDO under the 
Nanotechnology Materials Program ({\tt http://www.nedo.go.jp/kiban/nano/}),
and also partly by the Grant-in-Aid for ``Research for the
Future Program", Nano-carbons.

\section*{References}
$[1]$ Y. Shibayama, H. Sato, T. Enoki, and M. Endo, 
Phys. Rev. Lett. {\bf 84}, 1744 (2000).\\
$[2]$ K. Harigaya and T. Enoki, Chem. Phys. Lett. {\bf 351}, 128 (2002).\\
$[3]$ K. Harigaya, N. Kawatsu, and T. Enoki, 
in ``Nanonetwork Materials: Fullerenes, Nanotubes, and Related Systems",
(American Institute of Physics, 2001), pp. 529-532.
$[4]$ K. Harigaya, J. Phys.: Condens. Matter {\bf 13}, 1295 (2001).\\
$[5]$ K. Harigaya, Chem. Phys. Lett. {\bf 340}, 123 (2001).\\
$[6]$ A. M. Affoune, B. L. V. Prasad, H. Sato, T. Enoki, Y. Kaburagi, 
and Y. Hishiyama, Chem. Phys. Lett. {\bf 348}, 17 (2001).\\
$[7]$ Y. Kobayashi, K. Takai, J. Ravier, T. Enoki, and K. Harigaya, 
(unpublished results).\\
$[8]$ H. Ajiki and T. Ando, J. Phys. Soc. Jpn. {\bf 62}, 1255
(1993).\\
$[9]$ T. Ando and T. Nakanishi, J. Phys. Soc. Jpn. {\bf 67},
1704 (1998).\\
$[10]$ K. Harigaya, New J. Phys. {\bf 2}, 9 (2000).\\
$[11]$ K. Harigaya, J. Phys.: Condens. Matter {\bf 12}, 7069 (2000).\\
$[12]$ K. Kobayashi, Phys. Rev. B {\bf 53}, 11091 (1996).\\
$[13]$ T. M. Bernhardt, B. Kaiser, and K. Rademann,
Surface Science {\bf 408}, 86 (1998).\\

\end{document}